\documentclass[twocolumn,nofootinbib]{revtex4}
\usepackage{graphicx}
\usepackage{amsmath}
\usepackage{latexsym}
\usepackage[colorlinks, linkcolor=blue, citecolor=blue, urlcolor=blue]{hyperref}

\begin{document}

\title{Constraints on Dark Energy from New Observations including Pan-STARRS}

\author{Wei Zheng${}^{a,b,g}$}
\author{Si-Yu Li${}^{c}$}
\author{Hong Li${}^{b,d}$}
\author{Jun-Qing Xia${}^{b}$}
\author{Mingzhe Li${}^{e}$}
\author{Tan Lu${}^{f,g}$}

\affiliation{${}^a$Department of Physics, Nanjing University, Nanjing 210093, China}
\affiliation{${}^b$Key Laboratory of Particle Astrophysics, Institute of High Energy Physics, Chinese Academy of Science, P. O. Box 918-3, Beijing 100049, P. R. China}
\affiliation{${}^c$Theoretical Physics Division, Institute of High Energy Physics, Chinese Academy of Science, P. O. Box 918-4, Beijing 100049, P. R. China}
\affiliation{${}^d$National Astronomical Observatories, Chinese Academy of Sciences, Beijing 100012, P. R. China}
\affiliation{${}^e$Interdisciplinary Center for Theoretical Study, University of Science and Technology of China, Hefei, Anhui 230026, China}
\affiliation{${}^f$Purple Mountain Observatory, Chinese Academy of Sciences, Nanjing 210008, China}
\affiliation{${}^g$Joint Center for Particle, Nuclear Physics and Cosmology, Nanjing University -- Purple Mountain Observatory, Nanjing 210093, China}

\date{\today}

\begin{abstract}

In this paper, we set the new limits on the equation of state parameter (EoS) of dark energy with the observations of cosmic microwave background radiation (CMB) from Planck satellite, the type Ia supernovae from Pan-STARRS and the baryon acoustic oscillation (BAO). We consider two parametrization forms of EoS: a constant $w$ and time evolving $w(a)=w_0+w_a(1-a)$. The results show that with a constant EoS, $w=-1.141\pm{0.075}$ ($68\%~C.L.$), which is consistent with $\Lambda$CDM at about $2\sigma$ confidence level. For a time evolving $w(a)$ model, we get $w_0=-1.09^{+0.16}_{-0.18}$ ($1\sigma~C.L.$), $w_a=-0.34^{+0.87}_{-0.51}$ ($1\sigma~C.L.$), and in this case $\Lambda$CDM can be comparable with our observational data at $1\sigma$ confidence level. In order to do the parametrization independent analysis, additionally we adopt the so called principal component analysis (PCA) method, in which we divide redshift range into several bins and assume $w$ as a constant in each redshift bin (bin-w). In such bin-w scenario, we find that for most of the bins cosmological constant can be comparable with the data, however, there exists few bins which give $w$ deviating from $\Lambda$CDM at more than $2\sigma$ confidence level, which shows a weak hint for the time evolving behavior of dark energy. To further confirm this hint, we need more data with higher precision.

\end{abstract}


\maketitle


\section{Introduction}\label{Int}

Since the discovery of the accelerating expansion of the universe from the observations of type Ia supernovae (SNIa), dark energy, the most mysterious component in our universe which drives the acceleration, remains the hot topic in modern cosmology. The equation of state parameter (EoS), which is defined by the ratio of pressure and energy density, is one of the crucial parameters for characterizing the nature of dark energy. Therefore, extraction of the information on EoS plays important role in understanding dark energy.

Recently, many new data sets are released, such as, the first 15.5 months observational data from the Planck mission on the temperature power spectrum of CMB \cite{planck_fit}, and the new SNIa data from the Pan-STARRS observational project \cite{Panst1, Panst2}, which includes $112$ SNIa discovered in its first $1.5$ years along with a low-z sample of $201$ supernovae from other surveys and covers the redshift from $0.01$ to $0.63$. With the high quality data, the new limits are anticipated, for example, the Planck data favor more negative $w$ of dark energy and much lower value of Hubble constant compared with that obtained from previous WMAP observations \cite{planck, wmap9}, which may be a new hint for dark energy \cite{Xia:DE planck, quintom}. Also, with the accumulation of observational data sets, the better constraints on dark energy are anticipated, and in this paper we study the new constraints on EoS by performing the global data fitting analysis.

Fitting with the data, usually it needs introducing the parametrization for EoS, so that the evolution trajectory of dark energy component, such as the energy density, are described. The simplest model for EoS parametrization is $\Lambda$CDM, which assumes dark energy is cosmological constant and its EoS  $w=-1$. Although it is simple and can be comparable with the current datasets, however, many issues, such as the cosmological constant problems and the so called coincidence problems remain to be unanswered \cite{issue1, issue2, issue3}. Many other dark energy models, such as constant EoS and time evolving EoS assuming $w(a)= w_0+w_a(1-a)$, are also widely adopted in the literature \cite{CPL1, CPL2}. Although a parameterized EoS is simple and easily to be adopted in the data fitting analysis, it shapes the evolution of dark energy to be a special trajectory and this will cause bias. Comparing with the parameterized EoS, $w$ of the piecewise constant bins, which assumes EoS in certain redshift bins to be a constant instead of shaping $w$ in certain form, can avoid the bias introduced by the special parametrization and thus it is widely studied in recent years \cite{Christian, Gong-Bo-PRL109, Huterer and Turner, Paolo Serra, Yun Wang, Huterer_LPCA, PCA_ZGBO_08, Huterer_PCA, LPCA_ZGBO_10, Crittenden, Shi-Qi, YFW}.

Basing on the EoS of piecewise constant bins, the further model-independent studies, the \textit{principal component analysis} (PCA) method, are developed. The essence of PCA is to identify the direction of data points clustering in the parameter space and allows a dimension reduction of the parameter while keeping the loss of information minimum \cite{aa15281}. PCA was first adopted in addressing the issue of parametrization of dark energy properties by Huterer and Starkman in 2003 \cite{Huterer_PCA}, in which the $w(z)$ was expanded in terms of orthogonal function modes that were determined by the observational data. By truncating the poor modes, the reconstructed EoS can be less noise affected and more physical. Another variant type of PCA called local PCA was first taken by Huterer and Cooray in 2005 \cite{Huterer_LPCA}, which develops in recent years and is widely used in the dark energy model-independent data analysis \cite{PCA_ZGBO_08, LPCA_ZGBO_10, Christian, Shi-Qi, YFW, Najla}. This approach uses a set of new uncorrelated piecewise constant parameters, which can be regarded as localizing in the corresponding redshift regions, as the truly physical EoS of dark energy.

In this paper, we study the new constraints on dark energy by including the Planck temperature power spectra, low-$\ell$ WMAP9 polarization data, the SNIa sample released by the Pan-STARRS Project and the BAO measurements from Large Scale Structure (LSS) surveys. We consider the parameterized EoS, a constant EoS and the time-evolving EoS of $w(a)= w_0 + w_a(1-a)$, and the non-parameterized EoS with the PCA method. Our paper is organized as follows: In Section \ref{method} we describe the method for global fitting and the observational data sets used in the numerical analysis; Section \ref{result} and Section \ref{bin_eos} contains our mainly global constraints of the EoS for the parameterizations case and non-parameterized case from the current observations. The last Section \ref{summary} is the conclusions.


\section{Method and Data}\label{method}

\subsection{Numerical Method}

We perform a global fitting of cosmological parameters using the {\tt CosmoMC} package \cite{cosmomc}, a Markov Chain Monte Carlo (MCMC) code. We assume purely adiabatic initial conditions and a flat Universe. The pivot scale is set at $k_{s0} = 0.05{\rm Mpc}^{-1}$. The following basic cosmological parameters are allowed to vary with top-hat priors: the cold dark matter energy density parameter $\Omega_ch^2 \in [0.01, 0.99]$, the baryon energy density parameter $\Omega_bh^2 \in [0.005, 0.1]$, the scalar spectral index $n_s \in [0.5, 1.5]$, the primordial amplitude $\ln[10^{10}A_s] \in [2.7, 4.0]$, the ratio (multiplied by 100) of the sound horizon at decoupling to the angular diameter distance to the last scattering surface $100\Theta_s \in [0.5, 10]$, and the optical depth to reionization $\tau \in [0.01, 0.8]$. Besides these six basic cosmological parameters, we have introduced the EoS parameters for dark energy models. We consider the model of a constant EoS which introduces $w$ to remain constant ($c-w$ model), time evolving dark energy model which assumes $w(a)=w_0+w_a(1-a)$ ($te-w$ model) and we also consider EoS of piecewise constant bins ($bin-w$), in which we divide the redshift range into $N$ bins and assume EoS to be a constant in each bin. Then the parameters for dark energy can be summarized in the following equations,
\begin{equation}
w(z) = w_1 + \sum_{i=1}^{N-1}\frac{w_{i+1}-w_i}{2}\left[1+\tanh\left(\frac{z-z_{i+1}}{\xi}\right)\right]
\end{equation}
where $w_i$ stands for the value of EoS in the $i$th bin, $z_i$ and $z_{i+1}$ respectively denote the redshift of the start and end points of the $i$th bin, and the $\tanh$ function is adopted to link the neighbour two bins EoS transition smoothly. The parameter $\xi$ in the $\tanh$ function is used to control the transition width of $\tanh$, and in our calculation we control that such transition between two bins is sharp. This treatment guarantees that $w(z)$ can be handled as a smooth function, and the value of EoS in each bin can be approximately considered as a constant $w_i$. We have set the free parameters of $w$ in the scale of $w_i \in [-10,8]$ for each of dark energy models. When using the global fitting strategy to constrain the cosmological parameters, it is crucial to include dark energy perturbations \cite{xiapert}. In this paper we use the method provided in refs. \cite{xiapert,zhaopert} to treat the dark energy perturbations consistently in the whole parameter space in the numerical calculations, and we set the value of sound speed of dark energy perturbations $c_s^2 \equiv \delta p/ \delta \rho $ to unity. Therefore, the most general parameter space in the analyses is:
\begin{equation}
\{\Omega_{b}h^2, \Omega_{c}h^2, \Theta_{s}, \tau, n_s, A_s, w_{i}, f_{\nu}, N_{\rm eff}, \Omega_K\}~.
\end{equation}

\subsection{Current Observational Data}

In our analysis, we consider the following cosmological probes: i) temperature power spectra of CMB temperature from Planck satellite and low-$\ell$ polarization data from WMAP; ii) the baryon acoustic oscillation in the galaxy power spectra; iii) luminosity distances of type Ia supernovae.

For the Planck data from the 1-year data release \cite{planck_fit}, we use the low-$\ell$ and high-$\ell$ CMB temperature power spectrum data from Planck with the low-$\ell$ WMAP9 polarization data. We marginalize over the nuisance parameters that model the unresolved foregrounds with wide priors \cite{planck_likelihood}, and do not include the CMB lensing data from Planck \cite{planck_lens}.

Baryon Acoustic Oscillations, which measures the distance-redshift relation basing on the features in the clustering of galaxies of large scale surveys, provides an efficient method for measuring the expansion history. It measures not only the angular diameter distance, $D_A(z)$, but also the expansion rate of the universe, $H(z)$, which is powerful for studying dark energy \cite{task}. The traditional BAO data are not accurate enough for extracting the information of $D_A(z)$ and $H(z)$ separately \cite{okumura}, so one can only determine an effective distance \cite{baosdss}:
\begin{equation}
D_V(z)=[(1+z)^2D_A^2(z)cz/H(z)]^{1/3}~.
\end{equation}
Following the Planck analysis \cite{planck_fit}, in this paper we use the BAO measurement from the 6dF Galaxy Redshift Survey (6dFGRS) at a low redshift ($r_s/D_V (z = 0.106) = 0.336\pm0.015$) \cite{6dfgrs}, the measurement of the BAO scale based on a re-analysis of the Luminous Red Galaxies (LRG) sample from Sloan Digital Sky Survey (SDSS) Data Release 7 at the median redshift ($r_s/D_V (z = 0.35) = 0.1126\pm0.0022$) \cite{sdssdr7}, the BAO signal from BOSS CMASS DR9 data at ($r_s/D_V (z = 0.57) = 0.0732\pm0.0012$) \cite{sdssdr9}, and the BAO signal from WiggleZ measurement at $z=0.44, 0.60$ and $0.73$ \cite{wigg}. \footnote{In fact, during our preparation for the current paper, BAO measurement have obtained rapid progress, and many group provide the separated information of $D_A(z)$ and $H(z)$ thanks to the accumulation of large samples of galaxies \cite{bao_new1, bao_new2, bao_new3}, and we will consider to adopt them in our future study.}

For SNIa data, we adopt the first 1.5 years data release of Pan-STARRS project along with some other data from a combination of low-redshift surveys, which gives the Hubble diagram of $313$ SNIa and provides the most distant luminosity distance around redshift $0.6$. In the analysis, we take into account the systematic errors matrix provided by A. Rest and D. Scolnic \cite{Panst1, Panst2}. When calculating the likelihood, we marginalize over the absolute magnitude $M$, which is a nuisance parameter, as explained by Ref. \cite{planck_fit}. Comparing with Union2.1 \cite{Union2.1} or JLA (joint light-curve analysis) \cite{JLA}, the data set from Pan-STARRS first 1.5 years contains relatively smaller number of SNIa and lacks high-redshift supernovae (the maximum redshift is 0.63), which may weaken its constraining power in high redshift. However, the Pan-STARRS data has the advantage that its sample composition is relatively uniform, which means a more homogeneous distribution of supernova in redshifts and smaller systematic errors of distances modulus. Moreover, as a high-redshift program, Pan-STARRS survey is ongoing, so the shortage in SNIa total number and high redshift number will be overcome in the future, which is an important reason why we choose it as our SNIa data set for analysis.


\section{Numerical Results A}\label{result}

In this section we present the constraints on EoS parameters of dark energy from the global fitting analysis in two different scenarios: a constant EoS ($c-w$) and time-evolving EoS ($te-w$).

\subsection{Constraints on $c-w$ Scenario}

Firstly, we consider the constant EoS dark energy model. In table \ref{cw-table} we show the constraints on some related cosmological parameters from the data combinations of Planck, BAO and Pan-STARRS supernovae sample. By using the data combinations above, we obtain the $68\%$ constraint of $w=-1.141\pm{0.075}$, which is consistent with the previous work, such as the result $w=-1.149^{+0.078}_{-0.072}$ obtained by Rest et al. in ref. \cite{Panst2} and the one shown in Fig.1 of ref. \cite{Shafer}, both of which use the similar data combination Planck+BAO+Pan-STARRS like ours. These results show that the current compilation of Pan-STARRS is enough to provide good constraint on constant EoS of dark energy when combined with other probes. Similar tightly constraints on $w$ from Pan-STARRS can be compared with that from some other SNIa measurements. For example, the Planck+Pan-STARRS gives a $68\%$C.L. limit of $w=-1.173^{+0.084}_{-0.080}$ \cite{Panst2} obtained by Rest et al., which is slightly weaker than the constraints from Planck+JLA combination \cite{JLA} and WMAP7+Union2.1 combination \cite{Union2.1}. In fact, when using the SNIa data only, Pan-STARRS and Union2.1 give similar $68\%$C.L. constraint on $w$ of $w=-1.015^{+0.319}_{-0.201}$ \cite{Panst2} and $w=-1.001^{+0.348}_{-0.398}$ \cite{Union2.1} respectively, which shows the Pan-STARRS data is even stronger. On the other hand, the $\Lambda$CDM model with $w = -1$ is still compatible with our constraint on $w$ under the data combination of Planck+BAO+Pan-STARRS at $2\,\sigma$ confidence level. In spite of such consistency, the combination data sets including Pan-STARRS still show an insignificant tendency to $w<-1$, while the data combination including Union2.1 or JLA data are more compatiable with $\Lambda$CDM \cite{JLA, Shafer}. Moreover, our result is close to that constraint from Planck, which gives $w=-1.12^{+0.13}_{-0.14}$ $(95\% ~C.L.)$ under the data combination of Planck+BAO+SNLS.

\begin{table*}{\footnotesize
\caption{68$\%$ limits on base $\Lambda$CDM and different dark energy models from data combination Planck+BAO+Pan-STARRS.}
\label{cw-table}
\begin{center}
\begin{tabular}{c|c|c|c}

\hline
\hline
       & $\Lambda$CDM & $c-w$ model & $te-w$ model\\

\hline

       $w_0$  &  $-$  &  $-1.141\pm{0.075}$  &  $-1.09^{+0.16}_{-0.18}$            \\
\hline
       $w_a$  &  $-$  &  $-$  &  $-0.34^{+0.87}_{-0.51}$            \\

\hline
\hline

       $\Omega_b h^2$     &  $0.02215\pm0.00025$  &  $0.02200\pm0.00025$  &  $0.02197^{+0.00027}_{-0.00026}$            \\
\hline
       $\Omega_c h^2$     &  $0.1182\pm0.0016$  &  $0.1206\pm0.0020$  &  $0.1214\pm{0.0025}$            \\
\hline
       $100\theta$     &  $1.04152^{+0.00055}_{-0.00053}$  &  $1.04119^{+0.00058}_{-0.00057}$  &  $1.04110^{+0.00062}_{-0.00064}$          \\
\hline
       $\tau$     &  $0.092^{+0.012}_{-0.014}$  &  $0.089^{+0.012}_{-0.014}$  &  $0.087^{+0.012}_{-0.014}$            \\
\hline
       $n_s$     &  $0.9642\pm{0.0055}$  &  $0.9586\pm{0.0060}$  &  $0.9570^{+0.0069}_{-0.0068}$            \\
\hline
       $\ln(10^{10} A_s)$     &  $3.090^{+0.024}_{-0.027}$  &  $3.089\pm{0.025}$  &  $3.087^{+0.024}_{-0.027}$            \\

\hline
\hline

       $\Omega_{\Lambda}$     &  $0.6953^{+0.0094}_{-0.0097}$  &  $0.716\pm{0.014}$  &  $0.716\pm{0.015}$            \\
\hline
       $\Omega_m$     &  $0.3047^{+0.0097}_{-0.0094}$  &  $0.284\pm{0.014}$  &  $0.284\pm{0.015}$            \\
\hline
       $H_0~(\rm km\,s^{-1}\,Mpc^{-1})$     &  $68.04^{+0.72}_{-0.74}$  &  $71.1\pm{1.8}$  &  $71.3^{+1.8}_{-2.0}$            \\
\hline
       $Age~(Gyr)$     &  $13.793\pm{0.037}$  &  $13.752^{+0.040}_{-0.041}$  &  $13.739^{+0.049}_{-0.057}$            \\

\hline
\end{tabular}
\end{center}}
\end{table*}

\begin{figure}
\begin{center}
\includegraphics[scale=0.42]{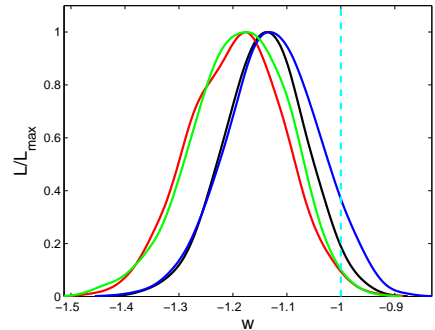}
\caption{One-dimensional marginalized distribution of $w$ under different models from data combination Planck+BAO+Pan-STARRS. The black line corresponds standard $c-w$ model, the red line corresponds $c-w$$\oplus$$\Omega_K$ model, the blue line corresponds $c-w$$\oplus$$N_{eff}$ model and the green line corresponds $c-w$$\oplus$$\sum m_{\nu}$ model. The vertical dashed line shows the location of value $w=-1$.}\label{lcdm}
\end{center}
\end{figure}

The constraint on $w$ is model-dependent. As shown in figure \ref{lcdm}, the constraints on $w$ could be changed in some extended $c-w$ dark energy models, due to the possible degeneracies between $w$ and other extended parameters in different models. In the model of allowing a non-zero spatial curvature, we get $w=-1.193^{+0.087}_{-0.088}$ ($1\,\sigma$ C.L.) and $\Omega_K=-0.0038\pm{0.0033}$ ($1\,\sigma$ C.L.); In the model of considering free effective number of neutrinos, we get $w=-1.126^{+0.086}_{-0.084}$ ($1\,\sigma$ C.L.) and $N_{eff}=3.23^{+0.31}_{-0.34}$ ($1\,\sigma$ C.L.); In the model of considering massive neutrinos, we obtain $w=-1.187^{+0.099}_{-0.081}$ ($1\,\sigma$ C.L.) and $\sum m_{\nu}<0.44eV~(95\% ~C.L.)$. Comparing with the results under standard $c-w$ model, the constraints on $w$ shown here under these extended $c-w$ models are all obviously weakened. Moreover, as the existence of correlations between $w$ and these extended parameters, data combinations prefer a more negative results on $w$ under some extended models. Under the framework of $c-w$$\oplus$$\Omega_K$, the coefficient of the correlation between EoS and spatial curvature is $cov(w,\Omega_K)=0.43$, so the slight tendency to a closed universe shown in the fitting value of $\Omega_K$ needs to be compensated by the more negative EoS as listed above, which rules out $w=-1$ at more than 2$\sigma$ C.L.. Similarly, under the model of $c-w$$\oplus$$\sum$$m_{\nu}$, the negative correlations between $w$ and $\sum m_{\nu}$  ($cov(w,\sum m_{\nu})=-0.53$) and a nonzero neutrino mass bring down the value of $w$, which also shows a deviation from $\Lambda$CDM at nearly 2$\sigma$ C.L..

\subsection{Constraints on $te-w$ Model}

\begin{figure}
\begin{center}
\includegraphics[scale=0.42]{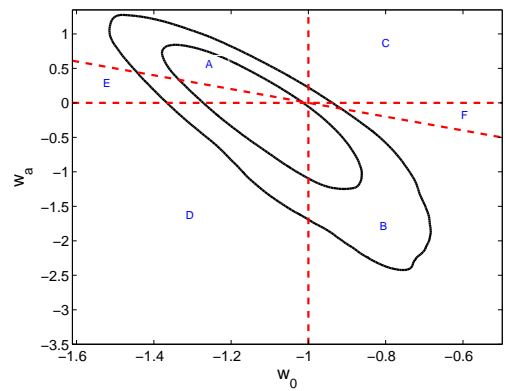}
\caption{Two-dimensional marginalized distribution between $w_0$ and $w_a$ from data combination Planck+BAO+Pan-STARRS. The three red dashed lines stand for $w_0=-1$, $w_0+w_a=-1$ and $w_a=0$, respectively. The capital letters stand for different dark energy models. See text for details.}\label{2d_w0wa}
\end{center}
\end{figure}

We present the constraints on time evolving dark energy model in which the EoS of dark energy is parameterized as $w(a)= w_0 + w_a(1-a)$ in this subsection. We summarize the numerical constraints on EoS parameters as well as some other cosmological parameters in table \ref{cw-table}. The constraints on EoS are $w_0=-1.09^{+0.16}_{-0.18}$ and $w_a=-0.34^{+0.87}_{-0.51}$ for $68\%~ C.L.$, which are consistent with our previous results from fitting with Planck and Union SNIa sample \cite{Xia:DE planck, Xia:2008ex, Li:2008vf, Li:2012ug, Li:2011, ZW:hz} and the Planck collaboration \cite{planck_fit}. In figure \ref{2d_w0wa} we show the two-dimensional constraints in the ($w_0$,$w_a$) panel which can be compared with the prediction of the $\Lambda$CDM model. We find that the $\Lambda$CDM ($w_0 = -1$, $w_a = 0$) is still favored at about $1\,\sigma$ confidence level. We have also separated the parameter space into six regions: two regions of $w(a)>-1$ (regions C and F for Quintessence dark energy models), two regions of $w(a)<-1$ (regions E and D for Phantom dark energy models), and two regions representing two kinds of Quintom models (A and B for Quintom model A and Quintom model B) of which $w(a)$ can cross $-1$ during its evolution, by the lines of $w_0=-1$, $w_0+w_a=-1$, and $w_a=0$. Note that the region E and F are two special Phantom and Quintessence regions respectively. For the region E, the corresponding EoS will keep below $-1$ even in the far future. Correspondingly, EoS in region F will remain $w>-1$ for ever. With the data selected contour, it gives that Qunitom dark energy models cover large area of parameter space. While the best fit point ($w_0=-1.09, w_a=-0.34$) is obviously located in the region D, which belongs to the Phantom dark energy models and implies the EoS crossing $-1$ in the future.

As can been seen that $w_0$ and $w_a$ are correlated with each other, which is doomed by the formation of the parametrization. To get uncorrelated EoS parameters in the 2-parameter framework, we can find a pivot redshift $z_p$ where $w$ is uncorrelated with $w_a$ and use $w(z_p)$ as a parameter instead $w_0$ to redo the parametrization \cite{pivot z_p}. But such parametrization is still model dependent. In the following, we will abandon the parametrization and discuss the model independent data fitting analysis of EoS parameters.


\section{NUMERICAL RESULTS B: bin eos} \label{bin_eos}

In this section, we do the model independent data fitting for dark energy. Abandoning the parametrization for EoS, we divide the redshift range into several bins and assume $w$ to be a constant in each bin. Considering that the supernovae of Pan-STARRS sample are mostly in the redshift range $z\in(0,0.6)$, most of our redshift bins for the EoS are set in (0, 0.6), and in each bin the constraining power from the data are almost comparable. For the choice of the number of bins, there is a balance we keep in mind. On the one hand, more bins can provide a easier way for us to trace the time evolving behavior of dark energy and reduce the parametrization dependence. On the other hand, you can not expect the number of bins to be infinity since the constraining power from the observational data is limited. In order to choose a suitable number of the bins, we introduce the $\chi^2$ statistic to pick out the bins's number.

By comparing the $\chi^2$ of the data fitting with picking different number of bins, we pick the 7 bins as a suitable choice for bin EoS analysis since it can give the smallest $\chi^2$. We set the $7$ bins localized between $z=0$ to $z=1$, in each bin the EoS $w$ is a free parameter, and when redshift goes beyond $1$, we fix $w=-1$, since the choice of the bins depends on the redshift of SNIa data sample. The nodes of the redshift bins are: $0$, $0.05$, $0.1$, $0.2$, $0.3$, $0.4$, $0.6$ and $1$. Fitting with the data sets of Planck, Pan-STARRS SNIa sample and BAO, we can get the constraints on $w_i$ in each bin.

\subsection{Constraints on w in Each Bin}
\begin{figure}
\begin{center}
\includegraphics[scale=0.43]{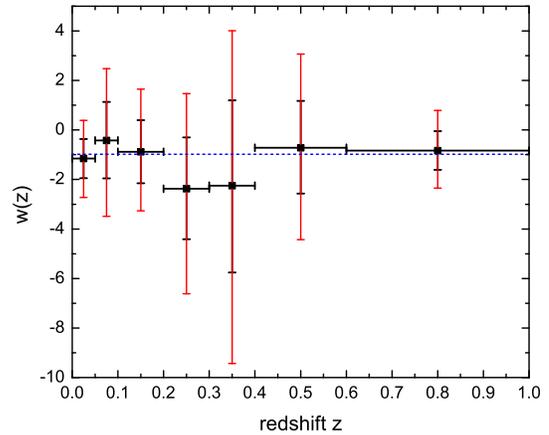}
\caption{The global fitting results about $bin-w$ from data combination Planck+BAO+Pan-STARRS. The inner(black) and outer(red) vertical error bars correspond to 68$\%$ and 95$\%$ error bars, and the horizontal error bars represent the range of each redshift bin.}\label{wi-mode}
\end{center}
\end{figure}

In figure \ref{wi-mode}, we show the constraints for $w_i$ after marginalizing over other cosmological parameters. The points with the error bars are located in the center of each redshift bins. Taking into account the corresponding error bars, we find that most of the mean values of these points are consistent with the prediction of $\Lambda$CDM model.

However, the EoS parameters $w_i$ are correlated, as the correlation coefficients between $w_i$ are usually not equal to $0$ from fitting with the data sets. The uncorrelated EoS are anticipated and are thought as the real description for dark energy. On the other hand, one can find that most of EoS parameters in different bins have relatively large errors, which means that the fitting result may be noise dominant and need some special treatment for excluding the noise and picking up the physical signal of EoS. In order to do that, we adopt the PCA method, and do the decorrelation treatment analysis. We also adopt the so called local PCA method for giving the EoS information in the local bins and we will introduce these results in the following two subsections.

\subsection{Principal Component Analysis}
In order to get uncorrelated information of EoS of dark energy, we adopt the PCA method \cite{Huterer_LPCA, Huterer_PCA}. From MCMC, we can compute the covariance matrix of $w_i$ by marginalizing over the other cosmological parameters,
\begin{equation}
C=<(w_i-\langle w_i\rangle)(w_j-\langle w_j\rangle)^T>=\langle \vec{p}\vec{p}^T\rangle-\langle \vec{p}\rangle \langle \vec{p}^T\rangle,
\end{equation}
where $\vec{p}$ is the vector of EoS parameters of dark energy $w_i$ and $\vec{p}^T$ is its transpose, and the Fisher matrix of $\vec{p}$ is $F=C^{-1}$. In order to get uncorrelated $w_i$, we should rotate $\vec{p}$ into a basis where the covariance matrix (or the Fisher matrix) is diagonal. To do that, we rotate the Fisher matrix by an orthogonal matrix $W$,
\begin{equation}
F=W^TDW,
\end{equation}
where $D$ is diagonal. The new parameters, now, can be written as $\vec{q}=W\vec{p}$ which are uncorrelated with each other because they have the diagonal covariance matrix $D^{-1}$. The $q_i$ are supposed to be the principal components (PCs) and the rows of the decorrelated matrix W, $e_i(z)$, are the eigenvectors (or the weights) which define the relations between the original parameters and the principal components.

There are many matrixes that can realize the diagonalization of $F$. The special type of decorrelated matrix which absorbs the diagonal elements of $D^{\frac{1}{2}}$ into the rows of W mentioned above, multiplying any orthogonal matrix $O$, $W^{*}=OD^{\frac{1}{2}}W$, can also diagonalize $F$ and make the parameters $q$ uncorrelated. In order to get uncorrelated EoS of dark energy which are physical without artificial treatment, we choose to adopt the following two kinds of realization: I. Normal principal component analysis: diagonalizing $F$ by an orthogonal matrix $W$, and then we order the eigenvalues of the diagonal matrix from small to large, by doing this we can fix the form of the orthogonal $W$ matrix. In this case, we can filter out the better constrained eigenmodes as well as the nosing modes. With the better constrained eigenmodes, we can reconstruct EoS of dark energy, which should be better constrained. II. We do the local PCA by choosing the decorrelated matrix $\widetilde{W}=F^{1/2}\equiv W^{T}D^{\frac{1}{2}}W$, and normalize $\widetilde{W}$ by making its rows sum to unity, which can ensure $q(z)=-1$ standing for $\Lambda$CDM . This choice has the advantage that the weights of $w_i$ are almost positive defined and fairly well localized in the redshift bins.

\subsubsection{weight I: normal PCA}
We diagonalize $F$, and realize $F=W^TDW$. The diagonal elements of $D$ are $d_i$, and each of uncorrelated parameters $q_i$ has the error $\sigma(q_i)=d_i^{-1/2}$. We order $d_i$ so that $\sigma(q_1)<\sigma(q_2)<....<\sigma(q_N)$. There are many orthogonal matrixes that can realize the diagonalization of F, however, after ordering the diagonal element of $D$, the decorrelated matrix $W$ is fixed to the only one.

\begin{figure}
\begin{center}
\includegraphics[scale=0.45]{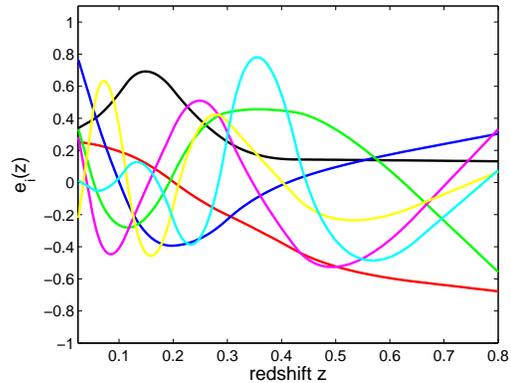}
\caption{The eigenvectors (or the weights) from different eigenmodes which correspond to the different PCs. The black line corresponds to $q_1$, the red line corresponds to $q_2$, the blue line corresponds to $q_3$, the green line corresponds to $q_4$, the magenta line corresponds to $q_5$, the yellow line corresponds to $q_6$ and the cyan line corresponds to $q_7$. }\label{emodes}
\end{center}
\end{figure}

We plot the eigenvectors of different modes in figure \ref{emodes}. From the shapes of the eigenvectors respect to redshift $z$, we can judge the better constrained modes, which have smaller error bar and less oscillation. While the eigenvectors which oscillate frequently and correspond to the eigenmodes having larger errors, are usually noise dominant, and can be ignored in principle. By keeping those good eigenmodes, we can reconstruct EoS $w(z)$ as:
\begin{equation}\label{eq7}
w(z)=\sum_{i=1}^{M} q_{i} e_i(z),
\end{equation}
where $M$ stands for the number of eigenmodes we adopt to reconstruct $w(z)$.

\begin{figure}
\begin{center}
\includegraphics[scale=0.44]{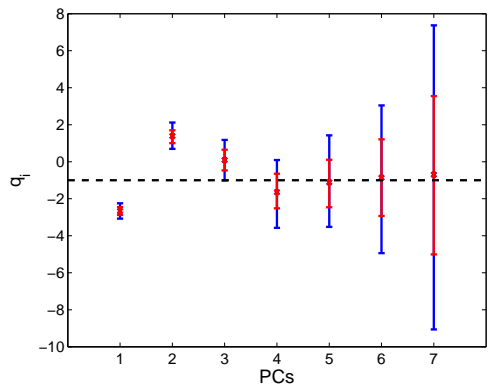}
\caption{The uncorrelated PCs (principal components) parameters $q_i$ denoted by PCs index $i$. The inner(red) and outer(blue) error bars correspond to 68$\%$ and 95$\%$ error bars, respectively.}\label{qi}
\end{center}
\end{figure}

\begin{figure}
\begin{center}
\includegraphics[scale=0.50,bb=25 0 612 473]{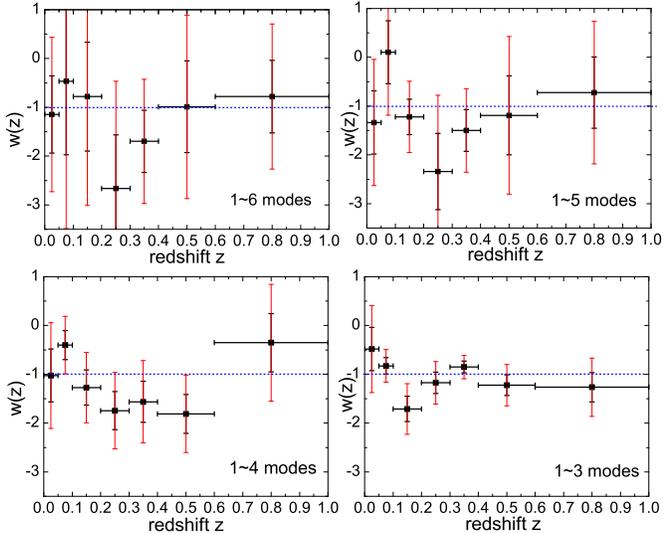}
\caption{The reconstructed $w(z)$ using the different number of eigenmodes. The inner(black) and outer(red) error bars correspond to 68$\%$ and 95$\%$ error bars, and the horizontal error bars represent the range of each redshift bin.}\label{df-em}
\end{center}
\end{figure}

We plot the uncorrelated parameters $q_i$ in figure \ref{qi}. Basing on these components, we reconstruct EoS parameters. In order to get better constraints in the reconstructed $w(z)$, we truncate the badly constrained eigenmodes and just consider the contribution from the good ones. By using Eq.(6), we compare four cases for adopting different number of eigenmodes respectively in figure \ref{df-em}. For adopting all the eigenmodes, it will recover the same $w(z)$ in figure \ref{wi-mode}.

\begin{figure}
\begin{center}
\includegraphics[scale=0.48]{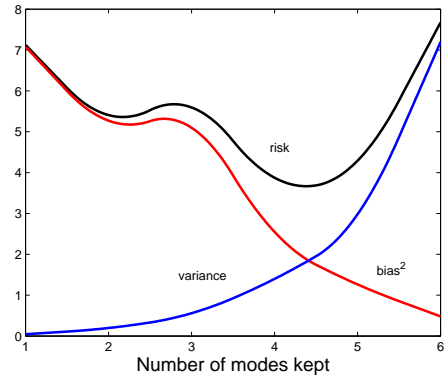}
\caption{Illustration of $risk$, $bias^2$ and $variance$ under including different number of eigenmodes. The black line represents the values of $risk$, the red line stands for the values of $bias^2$, the blue line stands for the values of $variance$.}\label{risk}
\end{center}
\end{figure}

In fact, when reconstructing $w(z)$, two things should be balanced properly, a). Adopting fewer modes into the reconstruction in order to avoid too much noise, however, it might lead to obvious bias; b). Adopting too much modes to avoid the risk of deviating from the true EoS, but this brings too much noise which will weaken the final constraints on EoS. To quantify such balance, we perform the calculation of $risk$ following the paper \cite{Huterer_PCA}, where
\begin{equation}\begin{split}
risk & =bias^2 + variance \\
     & =\sum_{i=1}^{N} [w(z_i)-{\bar{w}}(z_i)]^2+\sum_{i=1}^{N}{\sigma}^2(w(z_i))
\end{split}\end{equation}
Here, the $w(z_i)$ stands for the value of reconstructed $w(z)$ by taking into account different number of eigenmodes at the redshift $z_i$, and $\sigma(w(z_i))$ is its corresponding uncertainties. The $\bar{w}(z_i)$ denotes the fiducial value of $w(z)$ at redshift $z_i$, which can be approximately considered as the value of original bin-w. $N$ denotes the total number of bins. Thus, by using the Eq.(6), the $risk$ can be regarded as the function about number of eigenmodes to be kept, $M$. In the figure \ref{risk}, we illustrate the $risk$ value of considering different number of eigenmodes. Obviously, $M=4$ is the best choice to minimize the $risk$. With 4 eigenmodes, the reconstructed EoS are showed in the lower panel (left) of figure \ref{df-em}. Basing on this result, we find that most values of EoS are consistent with $\Lambda$CDM. However, there are still some bins that show slight deviation from $w=-1$, such as the second, fourth and sixth bins, the deviation is at about $95\% ~C.L.$. The EoS in the redshift $0.05<z<0.1$ behaves greater than $-1$, while in the redshift $0.2<z<0.3$ tends lower than $-1$, which imply a weak behavior of time-evolving dark energy with $w$ crossing $-1$. For using more eigenmodes to reconstruct $w(z)$, the larger error will obscure such behavior (upper two panels of figure \ref{df-em}) and with less eigenmodes, the oscillating hint is still obvious, as shown in the lower panel (right) of figure \ref{df-em}.

\subsubsection{weight II: local PCA}
Absorbing the diagonal elements of $D^{1/2}$ into orthogonal $W$, and multiplying another orthogonal matrix, is another useful realization, where we adopt $\widetilde{W}$, which is $\widetilde{W}\equiv W^{T}D^{1/2}W$, as the decorrelated matrix. The advantage of this choice is that the eigenmodes are localized distributed in each redshift bin and the weight of each mode is positive, and this kind of choice is considered as a useful basis for achieving the uncorrelated quantities, and is widely used in the analysis of the uncorrelated galaxies power spectrum \cite{Tegmark_galaxy_spectrum} and EoS of dark energy \cite{Huterer_LPCA, PCA_ZGBO_08, LPCA_ZGBO_10, Christian}.

\begin{figure}
\begin{center}
\includegraphics[height=150pt, width=260pt, scale=0.55, bb=40 0 612 473]{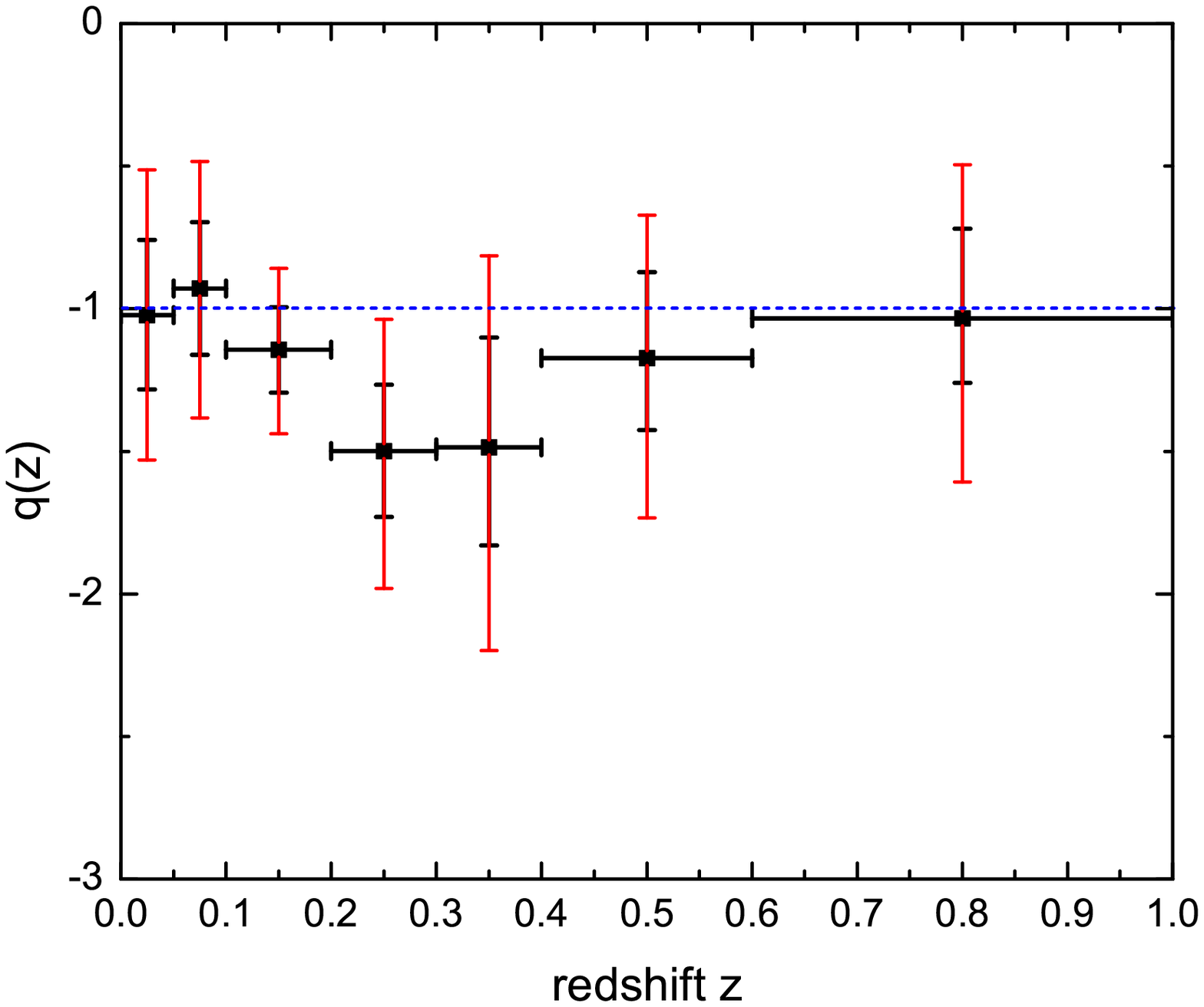}
\includegraphics[scale=0.42]{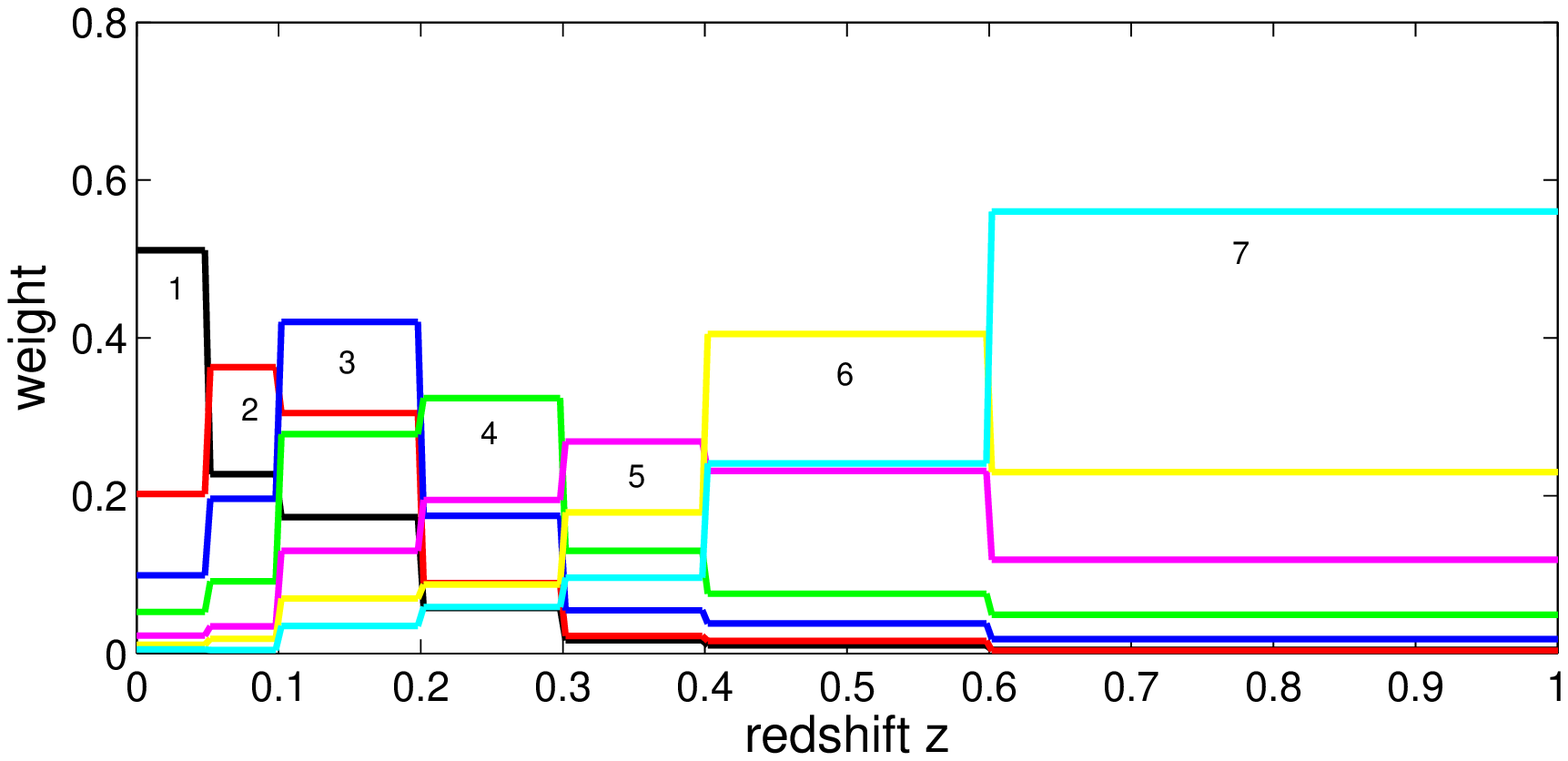}
\caption{The reconstructed EoS $q(z)$ (upper) and the weight functions (lower). In the upper panel, the inner(black) and outer(red) error bars correspond to 68$\%$ and 95$\%$ error bars, and horizontal error bars represent the range of each redshift bin. In the lower panel, the black line shows the weight function of $q_1$, the red line shows the weight function of $q_2$, the blue line shows the weight function of $q_3$, the green line shows the weight function of $q_4$, the magenta line shows the weight function of $q_5$, the yellow line shows the weight function of $q_6$, and the cyan line shows the weight function of $q_7$.}\label{lpca_qi_weight}
\end{center}
\end{figure}

In the upper panel of figure \ref{lpca_qi_weight}, we show the final $68\%$ and $95\%$ C.L. constraints on the seven uncorrelated parameters $q(z)$, which are thought as the uncorrelated EoS. We also plot the weight of each mode that does the decorrelation in the lower panel of figure \ref{lpca_qi_weight}. As shown in Fig. \ref{lpca_qi_weight}, most bins of $q(z)$ are consistent with $w=-1$ at $95\%$ C.L., and there exists weak hint for dynamical behavior of dark energy in the redshift around $0.1<z<0.6$, especially around the range of 4$th$ bin, which shows deviations from $w=-1$. However, considering the large errors, we need more accurate data to confirm it. In the lower panel of Fig. \ref{lpca_qi_weight}, we plot the weight for each bin, and the weight function well shows its positive and localized properties, which make the uncorrelated $q_i$ approximately one-to-one corresponds to original $w_i$ and better represents the true EoS.

\section{Summary}\label{summary}
Extracting the information of EoS plays crucial important role for understanding dark energy. In this paper, we discuss the latest constraints on EoS of dark energy with Planck, Pan-STARRS SNIa sample and BAO. We have performed 3 different data fitting by adopting a constant EoS, a time-evolving EoS and model independent method in which we divide redshift from $0$ to $1$ into several bins and assume EoS to be a constant in each bin.

When assuming $w$ to be a constant, we get $w=-1.14^{+0.14}_{-0.15}$ at $2\sigma~C.L.$, while fitting with time-evolving $w$, it gives that $w_0=-1.09^{+0.33}_{-0.30}$ and $w_a=-0.3^{+1.3}_{-1.5}$ at $2\sigma~C.L.$. Basing on those results, we can say that, at current stage, $\Lambda$CDM can be comparable with the observational data at about $2\sigma$ confidence level. For the $c-w$ model, the constraint on $w$ can be affected when allowing for different extended $c-w$ model parameters. For $te-w$ model, most area of confidence region in the $w_0$, $w_a$ parameter space is covered by the Quintom models but it is still consistent with $\Lambda$CDM model.  For the bin-w case, although the global fitting results about $w_i$ in each bin are all well consistent with $w=-1$, however, there is a weak hint in the model-independent PCA analysis that $w$ may deviate from $-1$ in some of the bins. Basing on the normal PCA, results about the reconstructed $w(z)$ imply a behavior of crossing $-1$ when ignoring severe badly constrained eigenmodes, while adding in more modes will enlarge the error bar and obscure this hint. The local PCA results about effective EoS $q(z)$ indicate a evidence of $w<-1$ in the $4$th bin but not very significant. To confirm such hint we should still keep in mind that the error bars are large for most of the bins.

To do further model independent data analysis of dark energy, we need more bins, still it need more SNIa. With the future progress in observations, hopefully, we can detect the signatures of the dynamics of dark energy or confirm whether or not EoS remains to be the cosmological constant.


\section*{Acknowledgements}

We acknowledge the use of the Legacy Archive for Microwave Background Data Analysis (LAMBDA). Support for LAMBDA is provided by the NASA Office of Space Science. HL is supported in part by the National Science Foundation of China under Grant Nos. 11033005, by the 973 program under Grant No. 2010CB83300, by the Chinese Academy of Science under Grant No. KJCX2-EW-W01. JX is supported by the National Youth Thousand Talents Program. ML is supported by Program for New Century Excellent Talents in University and by NSFC under Grants No. 11075074. TL is supported by the National Science Foundation of China under Grant Nos. 11373068 and Nos. 10973039. The research is also supported by the Strategic Priority Research Program ¡°The Emergence of Cosmological Structures¡± of the Chinese Academy of Sciences, Grant No. XDB09000000.


\end{document}